\documentclass[11pt,onecolumn]{article}

\usepackage{amsmath,amssymb}
\usepackage{bm}
\usepackage{upgreek}
\usepackage{xcolor}
\usepackage{graphicx}
\usepackage{caption}
\usepackage{subcaption}
\usepackage{tabularx}
\usepackage{enumitem}
\usepackage[colorlinks=true,
            linkcolor=blue,
            citecolor=blue,
            urlcolor=blue]{hyperref}

\usepackage[a4paper,margin=1.2in]{geometry}

\newcommand{\tr}{\text{tr}}

\newcommand{\bb}[1]{\bm{#1}}
\newcommand{\cb}[1]{\bb{\mathcal{#1}}}

\newcommand{\Lcb}{\ensuremath{\cb{L}}}
\newcommand{\Nb}{\ensuremath{\bb{N}}}
\newcommand{\hb}{\ensuremath{\bb{h}}}
\newcommand{\dodo}{\mathbin{:}}
\newcommand{\bs}[1]{\boldsymbol{#1}}

\begin{document}


\title{Mechanical and Structural Contributions to Anisotropy in Granular Materials}

\author{
Mehdi Pouragha$^{1}$,
Gertraud Medicus$^{2}$,
Selvarajah Premnath$^{1}$,
Siva Sivathayalan$^{1}$
}

\date{
\small $^{1}$Carleton University, Ottawa, Canada\\
$^{2}$University of Innsbruck, Austria
}

\maketitle

\vspace*{0.5cm}

\noindent\rule{\textwidth}{0.5pt}
{\small {\bf Abstract}. Anisotropy in granular materials arises from both the internal fabric and the directionality of the stress state, yet separating these effects experimentally remains challenging. This study develops a first–order linearisation of the incremental stress–strain response that isolates mechanical anisotropy from structural anisotropy using two independent orientation measures. The formulation enables both contributions to be quantified directly from macroscopic laboratory data. The method is applied to hollow-cylinder tests with systematically varied loading directions. Results show that both anisotropy components intensify as the stress state becomes more deviatoric; mechanical anisotropy is consistently stronger; and its relative dominance decreases with increasing deviatoric stress. Comparison with an isotropic hypoplastic model confirms that mechanically induced directional effects are captured even without fabric anisotropy. The framework offers a practical and physically transparent means for quantifying and comparing anisotropy mechanisms in granular materials.

}

\noindent\rule{\textwidth}{0.5pt}

\section{INTRODUCTION}

The directional dependence of the mechanical response of soils is one of the most consistently observed characteristics of granular materials. Experiments using triaxial, plane--strain, and hollow--cylinder devices show that sand strength, stiffness, and deformation patterns depend strongly on loading orientation relative to the sample’s deposition or consolidation history \cite{arthur1980principal,symes1984undrained, fakharian2022influences,sivathayalan2002influence,wijewickreme1993behaviour,wijewickreme2008experimental,prasanna2020effects}. The resulting anisotropy affects both the mobilised strength and the non--coaxiality between stress and strain increments, and it plays a critical role in phenomena such as strain localisation and cyclic instability. Despite decades of research and advances in multiscale experiments, a unified framework for describing and quantifying soil anisotropy remains elusive.

Traditionally, soil anisotropy has been classified as \emph{inherent} or \emph{induced} \cite{wong1985induced,oda1993inherent}. The former is attributed to the depositional and geological history of the material, while the latter develops during loading through changes in particle arrangement and contact orientation. Although this dichotomy captures the qualitative origins of anisotropy, it offers limited value for describing a given material state: for any specimen under load, there is no objective means of quantifying the respective contributions of deposition history and subsequent deformation.

From a mechanics perspective, anisotropy may be viewed as a manifestation of directional symmetry breaking in the material response. Under full rotational symmetry, representation theorems allow the stress-strain relation to be reduced to scalar invariants, reflecting the absence of any preferred direction. This invariant formulation underlies most isotropic critical--state and elasto--plastic models, such as those based on the Cam-clay theory. Once this symmetry is broken, however, a description in terms of invariants alone constitutes only a zeroth--order approximation, the adequacy of which diminishes with increasing anisotropy. 

In the absence of a clear method to distinguish between different sources of anisotropy, directional dependencies observed in experiments or numerical simulations are often attributed primarily to internal material anisotropy, even though a break in directional symmetry need not originate from the fabric itself: a non--isotropic stress state already defines a preferred direction in an otherwise isotropic medium. In practice, stress-induced and fabric-induced anisotropies often coexist and cannot be readily distinguished in experiments.

Consequently, for defining the state of a soil under a given stress condition, the traditional inherent/induced distinction is of limited practical use. A more meaningful distinction arises by separating \emph{mechanical anisotropy}, associated with the anisotropy of the stress state, from \emph{structural} (or \emph{geometrical}) anisotropy, associated with the internal fabric of the material. This view aligns with micromechanical studies \cite{rothenburg1989analytical,zhao2013unique}, in which the anisotropy of stress is expressed in terms of the anisotropy of contact forces, while the structural anisotropy reflects the orientation distribution of contacts or particles.

Recent years have seen remarkable advances in directly measuring fabric anisotropy 
using micro-imaging techniques such as X-ray tomography and photoelasticity%
~\cite{wiebicke2020measuring,yang2008quantifying,imseeh20183d,oda1985stress,majmudar2005contact}. 
These developments complement indirect approaches based on calibrating 
anisotropic constitutive models to laboratory data~\cite{gu2024quantifying,kuwano2000anisotropic,symes1984undrained,li2012anisotropic,pouragha2020micromechanical}. 
Despite this progress, most quantitative insights into the evolution and effects of 
anisotropy still rely on discrete-element simulations, which provide direct access to contact orientations and force chains. 
As a result, experimental methods capable of distinguishing mechanical from structural anisotropy using only macroscopic response measurements remain rare.

The present study develops a rigorous first--order linearisation of the incremental stress--strain response that separates the contributions of mechanical and structural aniso\-tropy in granular materials. Starting from an objective rate formulation and applying representation theorems for isotropic tensor functions, the incremental response is expressed in terms of two independent measures of loading orientation: one associated with the current stress state, and the other with the internal fabric. These orientation measures make it possible to quantify separately how the incremental response is influenced by the stress state and by the internal fabric, thereby isolating the contributions of mechanical and structural anisotropy without invoking any specific constitutive model.

To apply this framework, the present study analyses a set of hollow--cylinder experiments in which the loading direction is varied systematically. The experimental programme of \cite{premnath2023effects} is particularly well suited to this purpose, because its two complementary loading configurations, one aligning the consolidation stress with the depositional fabric, and the other aligning it with the imposed incremental strain  direction, create controlled contrasts between mechanical and structural sources of anisotropy. By examining the initial undrained stress--path inclination across a range of loading orientations, the present study extracts quantitative measures of the two anisotropy components directly from macroscopic test data, without recourse to discrete--element simulations or microstructural observations. This makes the approach unique in its ability to quantitatively decompose the two mechanisms contributing to anisotropy using laboratory data alone. Application of this framework to the experimental results reveals several consistent trends. Both mechanical and structural anisotropy intensify as the stress state becomes more deviatoric; mechanical anisotropy remains the dominant contribution across all states examined; yet its relative dominance decreases as the deviatoric stress increases. 

Finally, the experimental observations are compared with predictions from an isotropic hypoplastic model. Although the model contains no representation of structural anisotropy, it reproduces a significant portion of the directional effects associated with mechanical anisotropy in the non-coaxial tests. The model predictions exhibit an almost linear dependence of the initial response on loading orientation, thereby providing further support for the validity of the proposed linearisation framework.

\section{Notation}

The following notation is used throughout this paper. 
Given a second--order tensor $\bs{X}$, Euclidean norm $\|\bs{X}\|$, and normalised unit tensor, $\widehat{\bs{X}}$ $=\bs{X}/\|\bs{X}\|$ are defined as
\begin{equation}\label{eq:additive_Aniso}
\begin{gathered}
    \|\bs{X}\| = \sqrt{\bs{X}\dodo\bs{X}}, \qquad \widehat{\bs{X}}=\bs{X}/\|\bs{X}\| .
\end{gathered}
\end{equation}

The mean and deviatoric components of the Cauchy stress tensor $\bs{\sigma}$ 
and infinitesimal strain tensor $\boldsymbol{\varepsilon}$ are written as
\begin{equation}
\begin{gathered}
 p = \tfrac{1}{3}\,\tr(\bs{\sigma}), 
    \qquad 
    \bs{s} = \bs{\sigma} - p\,\bm{I},
    \qquad 
    q = \sqrt{\tfrac{3}{2}\,\bs{s}\dodo\bs{s}}\\
    \varepsilon_v = \tr(\bs{\varepsilon}), 
    \qquad 
    \bs{e} = \bs{\varepsilon} - \tfrac{1}{3}\,\varepsilon_v\,\bm{I},
    \qquad 
    \varepsilon_q = \sqrt{\tfrac{2}{3}\,\bs{e}\dodo\bs{e}} .
\end{gathered}
\end{equation}
Here $\bm{I}$ denotes the second--order identity tensor, and the operator ``$\dodo$'' represents the double contraction 
$\boldsymbol{A}\dodo\boldsymbol{B} = \tr(\boldsymbol{A}^{\mathsf{T}}\boldsymbol{B})$. 
A superposed dot $(\dot{\;})$ indicates a time rate or incremental rate. All stress variables used in this paper refer to Terzaghi’s effective stress.

\section{Theoretical framework}

In general, the anisotropic response of a granular material can be expressed through a constitutive relation that links the stress rate to the strain rate and to the current state of stress and internal structure. In rate form, this can be written as
\begin{equation}\label{eq:GeneralConst}
    \dot{\bs{\sigma}} = \mathcal{G}(\bs{\sigma}, \bs{F}, \dot{\bs{\varepsilon}}),
\end{equation}
where $\bs{\sigma}$ and $\bs{\varepsilon}$ denote the Cauchy stress and infinitesimal strain tensors, respectively, and $\bs{F}$ is a generic traceless internal (fabric) tensor characterising the structural anisotropy of the material. In reality, the constitutive relation also depends on additional scalar state variables such as the void ratio, but these are omitted here, without loss of generality, for the sake of clarity in the subsequent arguments. Both $\bs{\sigma}$ and $\bs{F}$ define preferred directions and may therefore introduce directional dependence into the incremental response, corresponding to what we term \emph{mechanical} and \emph{structural} anisotropy.   The detailed form of $\mathcal{G}(\cdot)$ is not considered here; instead, attention is restricted to the quantitative influence and relative magnitude of mechanical and structural anisotropies through a linearised representation.

\subsection{Linearised representation}

At a given reference state, the incremental constitutive mapping in Eq.~\eqref{eq:GeneralConst} defines the rate of change of stress as an objective tensor function of the stress tensor, fabric tensor, and strain--rate tensor. 
According to Boehler’s representation theorem for isotropic tensor functions of several symmetric arguments, e.~g.~ 
\cite{boehler1979simple, Spencer1971}, such a function can be expressed in terms of the isotropic tensor basis generated by $\bs{\sigma}$, $\bs{F}$, and $\dot{\bs{\varepsilon}}$, with scalar coefficients that depend only on the joint invariants of these tensors. 
Following this theorem, the incremental response can be expanded in a hierarchy of invariants of increasing order. 
We retain only the lowest-order terms that are linear in the deviatoric strain--rate tensor, corresponding to a first-order directional expansion of the incremental response.  Focusing on the instantaneous stress-path inclination in the $(p,q)$-plane, the model-agnostic linearisation analysis presented in the Appendix shows that the slope $\dot{q}/\dot{p}$ may be locally approximated, about the aligned reference state $\Omega_\sigma=\Omega_F=1$, in terms of the two cosine-like anisotropy measures $\Omega_\sigma$ and $\Omega_F$ as 
\begin{equation}\label{eq:linPhi}
    \frac{\dot q}{\dot p}
    = \eta_0 + A_\sigma \,\Omega_\sigma + A_F\,\Omega_F,
\end{equation}
with
\begin{equation} \label{eq:Omegas}
\begin{split}
    \Omega_\sigma
      &:= \frac{\bs{s}\dodo\dot{\bs{e}}}{\|\bs{s}\|\,\|\dot{\bs{e}}\|}=\widehat{\bs{s}}\dodo \widehat{\dot{\bs{e}}},\quad
    \Omega_F
      := \frac{\bs{F}\dodo\dot{\bs{e}}}{\|\bs{F}\|\,\|\dot{\bs{e}}\|}=\widehat{\bs{F}}\dodo \widehat{\dot{\bs{e}}}.
\end{split}
\end{equation}
Here, the coefficients $\eta_0$, $A_\sigma$, and $A_F$ are scalar functions of the current state variables,  but are independent of the instantaneous loading direction, as derived in the Appendix.
The cosine-like invariants, $\Omega_\sigma$ and $\Omega_F$ quantify the degree of  non--coaxiality of the deviatoric strain rate with the deviatoric stress and fabric tensors, respectively. The  coefficients $A_\sigma$ and $A_F$ quantify the first-order sensitivity to the alignment of the strain-rate deviator with the stress deviator (mechanical anisotropy) and the fabric (structural anisotropy), respectively. At the aligned reference state, $\Omega_\sigma=\Omega_F=1$, the instantaneous slope is $\eta_0 + A_\sigma \, + A_F$. Crucially, the expression in Eq.~\eqref{eq:linPhi} allows the relative significance of the mechanical and structural anisotropies at a given state to be assessed by comparing $A_\sigma$ and $A_F$.

It is important to note that the coefficients $\eta_0$, $A_\sigma$, and $A_F$ are independent of the instantaneous strain--rate tensor and depend only on the current state. These coefficients are assumed constant for a given state, allowing them to be estimated from hollow-cylinder tests with comparable states but different loading orientations, as will be demonstrated next.

\subsection{Identification of $A_\sigma$ and $A_F$ through complementary loading paths}

The parameters $A_\sigma$ and $A_F$ in Eq.~\eqref{eq:linPhi} represent, respectively, the first--order sensitivities of the incremental response to mechanical and structural anisotropy. Their quantitative identification requires experimental conditions under which the effects of stress--induced and fabric--induced anisotropies can be observed separately. This concept is applied to two complementary series of hollow--cylinder tests reported by \cite{premnath2023effects}, which provide distinct loading conditions for isolating the two effects. The two series were originally referred to as \emph{Unmatched} and \emph{Matched}; for theoretical clarity, they are herein termed the \emph{non--coaxial} and \emph{stress--coaxial} loading paths, respectively. 

\begin{figure}
	\centering
	\includegraphics[width=0.6\textwidth]{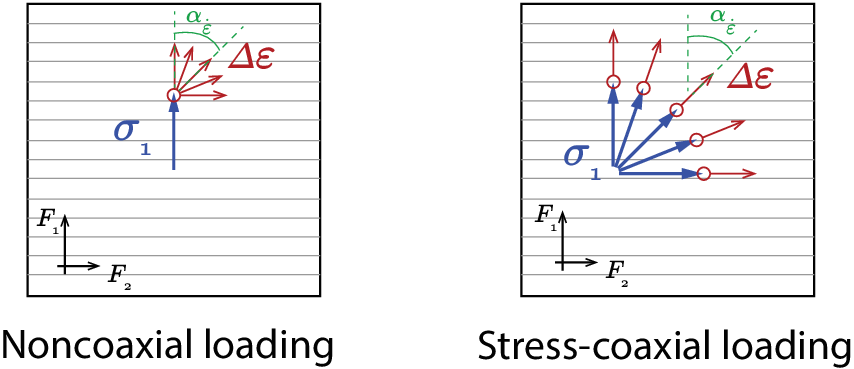}
	\caption{Schematics of consolidation stress and loading conditions for non--coaxial test series (left),  and stress--coaxial test series (right).}
	\label{fig:schematics}
\end{figure}

In the \emph{non--coaxial} series, the principal axes of the consolidation stress are aligned with the depositional fabric, while the subsequent undrained loading is applied with a rotated principal strain--rate direction, $\alpha_{\dot{\bm{\varepsilon}}}$, see Fig.~\ref{fig:schematics}-(top). The precise definition of $\alpha_{\dot{\bm{\varepsilon}}}$ is found in \cite{premnath2023effects}. This configuration produces an initial alignment between stress and fabric, but a misalignment between the stress and strain--rate tensors, thereby introducing directional dependence  through both the internal fabric and the stress state. The observed anisotropic response therefore originates from a combination of mechanical and structural anisotropy. 

In contrast, the \emph{stress--coaxial} series involves specimens consolidated along a stress direction that is already aligned with the subsequent loading path, see Fig.~\ref{fig:schematics}-(bottom). Here, stress and strain--rate remain coaxial at the onset of shearing, and the directional response arises predominantly from the internal structure of the material. This configuration thus highlights the contribution of structural anisotropy represented by the fabric tensor~$\bs{F}$, while the direct mechanical influence of stress anisotropy is minimised.

Both test series have been conducted at various initial deviatoric stress conditions described by $K_c=\sigma_1/\sigma_3\,\in\, \{1.5, 2.0, 2.5 \}$, and $b_\sigma = (\sigma_2-\sigma_3)/(\sigma_1-\sigma_3)=0.4$.
Recalling the definition of $\Omega_F$ and $\Omega_\sigma$, we can reasonably assume that the following conditions hold for the two test series
\begin{equation}
\begin{split}
    \text{non--coaxial series:} \quad \widehat{\bs{F}}\dodo\widehat{{\bs{s}}}&=1 \,\to\, \Omega_F=\Omega_\sigma,\\
    \text{stress--coaxial series:} \quad \widehat{\bs{s}}\dodo\widehat{\dot{\bs{e}}}&=1 \,\to\, \Omega_\sigma=1.
\end{split}
\end{equation}
Substituting these into Eq.~\eqref{eq:linPhi} gives:
\begin{equation}\label{eq:NC_SC_conditions}
\begin{split}
    \text{non--coaxial series:} \quad &\frac{\dot q}{\dot p}
    = \eta_0 + (A_\sigma \, + A_F)\,\Omega_\sigma,\\
    \text{stress--coaxial series:} \quad &\frac{\dot q}{\dot p}
    = \eta_0 + A_\sigma  + A_F\,\Omega_F.
\end{split}
\end{equation}
Interestingly, Eq.~\eqref{eq:NC_SC_conditions} implies that the coefficients $A_\sigma$ and $A_F$ can be estimated directly from the initial response of the two test series. The analysis procedure is as follows: 
\begin{enumerate}[leftmargin=1.2em,labelindent=0pt,itemindent=0pt, itemsep=0pt, parsep=0pt]
    \item Compute $\Omega_\sigma$ and $\Omega_F$ for non--coaxial and stress--coaxial paths, respectively, 
    \item Calculate the instantaneous  slope of the stress path, $({\dot q}/{\dot p})_\text{ini}$, at the beginning of undrained shearing for each loading direction, 
    \item Plot $({\dot q}/{\dot p})_\text{ini}$ vs $\Omega_\sigma$ or $\Omega_F$, for a given value of $K_c$, and use the slope to calculate $A_\sigma$ and $A_F$ from Eq.~\eqref{eq:NC_SC_conditions}, and
    \item Repeat the procedure for all values of $K_c$ to explore the variation of $A_\sigma$ and $A_F$, and their ratio with $K_c$.
\end{enumerate}

\section{Results and Discussion}

\begin{figure*}
	\centering
	\includegraphics[width=1\textwidth]{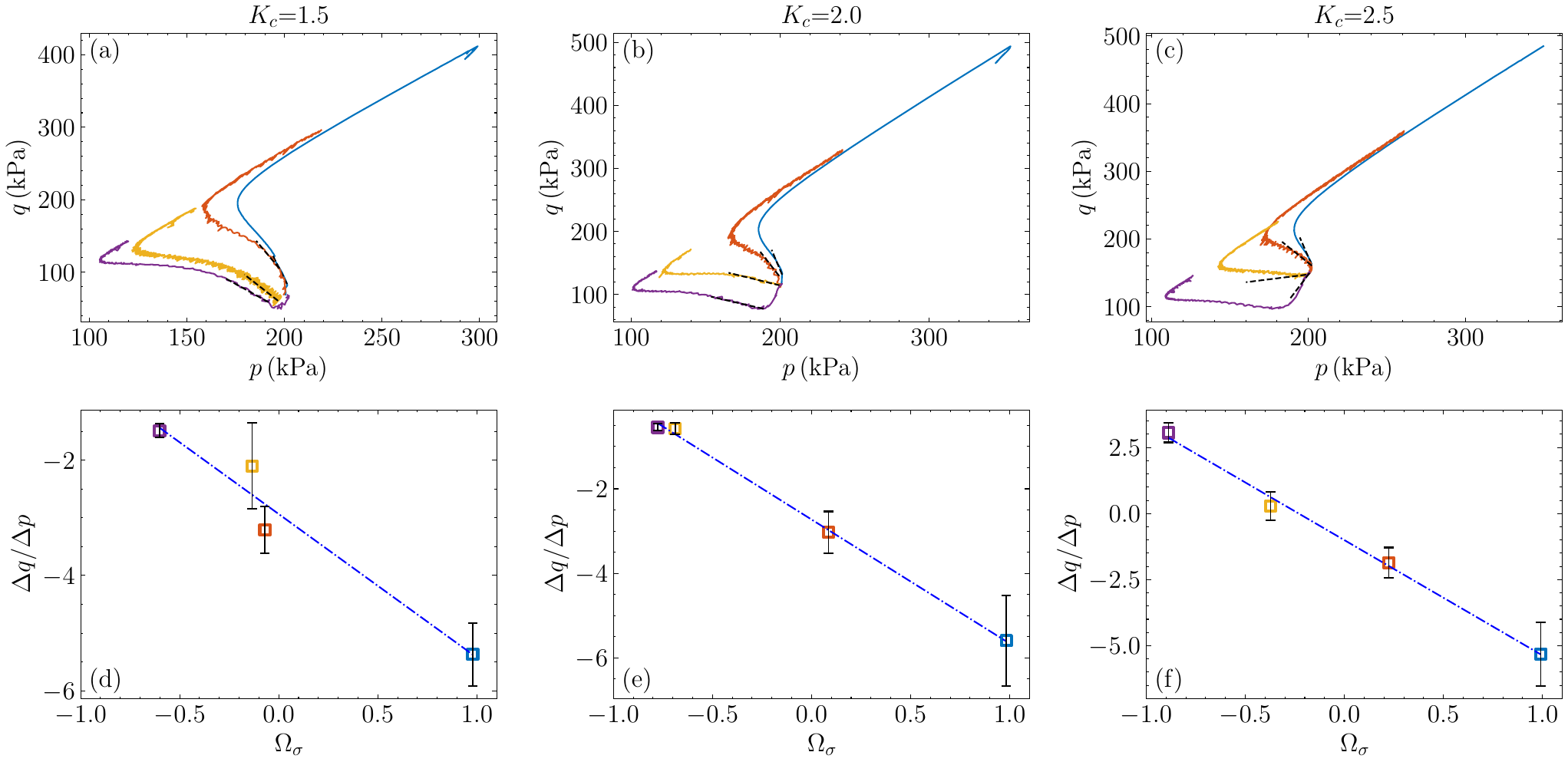}
	\caption{Effective stress paths (top row) for tests with various $\alpha_{\dot{\varepsilon}}$, and variation of $\frac{\dot{q}}{\dot{p}}$ with $\Omega_\sigma$ (bottom row) for \textit{non--coaxial loading} experiments with (a, d) $K_c = 1.5$, (b, e) $K_c = 2.0$, and (c, f) $K_c=2.5$.}
	\label{fig:nonX}
\end{figure*}

\begin{figure*}
	\centering
	\includegraphics[width=1\textwidth]{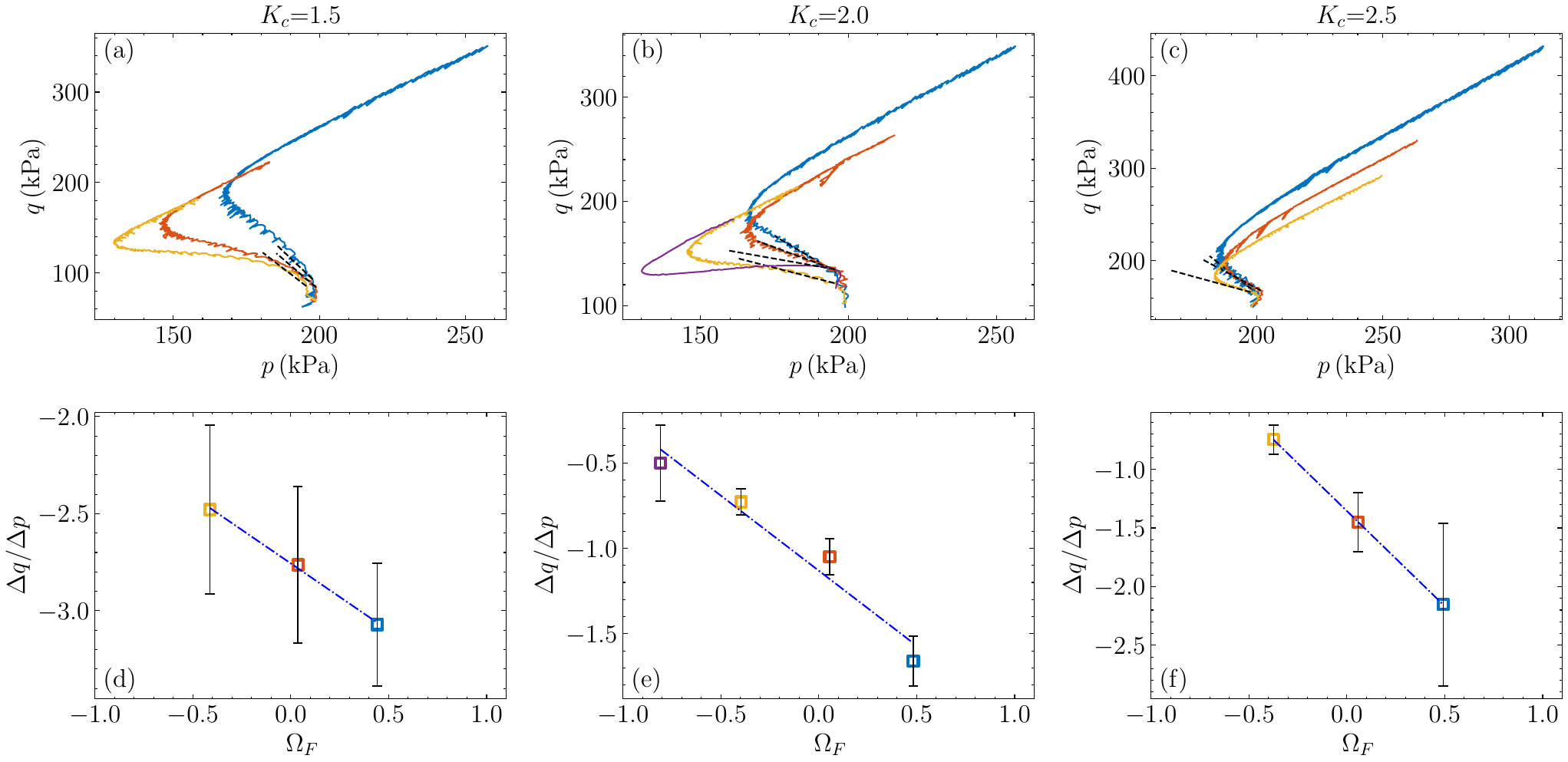}
	\caption{Effective stress paths (top row) for tests with various $\alpha_{\dot{\varepsilon}}$, and variation of $\frac{\dot{q}}{\dot{p}}$ with $\Omega_F$ (bottom row) for \textit{stress--coaxial loading} experiments with (a, d) $K_c = 1.5$, (b, e) $K_c = 2.0$, and (c, f) $K_c=2.5$.}
	\label{fig:coX}
\end{figure*}

This section presents the analysis of experimental data from the hollow--cylinder tests of \cite{premnath2023effects}, used to quantify the coefficients $A_\sigma$ and $A_F$ representing the mechanical and structural components of anisotropy. For each test, the instantaneous slope of the stress path, $(\dot q / \dot p)_{\text{ini}}$, was evaluated from the initial undrained loading segment. A brief non–constitutive transient, occasionally observed immediately after the change in loading direction, was excluded to avoid artefacts associated with controller settling. In addition, a small number of tests exhibiting pronounced initial unloading were excluded, as the stress state upon reloading no longer matched that of the other tests and therefore did not satisfy the requirement of comparable initial states. The subsequent portion of the $q$–$p$ path was used for the linear fit shown as black dashed lines in Figs.~\ref{fig:nonX}(a–c) and \ref{fig:coX}(a–c). To quantify the sensitivity of the fitted slope to the chosen interval, the fitting window was shifted by $\pm 15$ measurement points, and the standard deviation of the resulting slopes within that window was computed and shown as error bars in Figs.~\ref{fig:nonX}(d–f) and \ref{fig:coX}(d–f).

The bottom rows of Figs.~\ref{fig:nonX} and \ref{fig:coX} show the variation of the measured initial stress–path inclination with the orientation measures $\Omega_\sigma$ (non--coaxial series) and $\Omega_F$ (stress--coaxial series), respectively. In both loading configurations, a clear near–linear dependence is observed across all $K_c$ values, with modest scatter. Interpreting these trends using Eq.~\eqref{eq:NC_SC_conditions}, the non--coaxial series reflects the compounded influence of mechanical and structural anisotropies (slope proportional to $A_\sigma + A_F$), whereas the stress--coaxial series isolates the orientation dependence associated with the internal fabric (slope proportional to $A_F$). Consistently, the orientation dependence observed in the stress--coaxial series is smaller than that in the non--coaxial series, as expected when only one mechanism is active.

\begin{figure}
	\centering
	\includegraphics[width=0.95\textwidth]{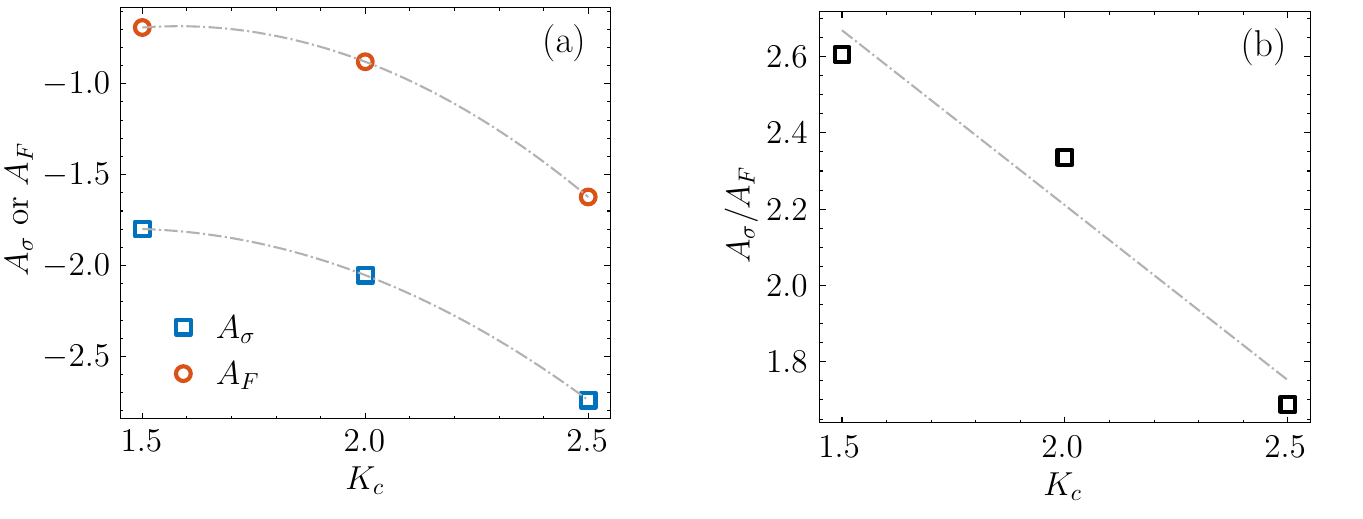}
	\caption{Variation with $K_c$ of; (a) $A_\sigma$ and $A_F$ parameters from Eq.~\eqref{eq:linPhi}, and (b) the ratio of $\frac{A_\sigma}{A_F}$.}
	\label{fig:BC_trend}
\end{figure}

The anisotropy coefficients extracted from the fitted trends are summarised in Fig.~\ref{fig:BC_trend}. Both coefficients become more negative with increasing $K_c$ (Fig.~\ref{fig:BC_trend}-(a)). For $A_\sigma$, this reflects the well-established effect that stress-induced anisotropy becomes more pronounced as the stress state departs from isotropy; a larger deviatoric stress amplifies the contribution of the cross-invariant $\boldsymbol{s}\dodo\dot{\boldsymbol{e}}$, consistent with hypoplasticity and generalised plasticity formulations in which directional response grows with shear stress. The increase in the magnitude of $A_F$ with $K_c$ does not imply a change in the fabric itself, which is assumed to remain fixed throughout the test series. Rather, it reflects the greater relative sensitivity of the stress ratio $q/p$ to loading direction when the stress state is more deviatoric. This observation is akin to a larger logarithmic ratio $(\dot q/q)/(\dot p/p)$ which enhances the influence that a fixed fabric orientation exerts on the incremental response. In this picture, the observed increase of $A_F$ with $K_c$ arises from a state-dependent amplification of the fabric’s effect, not from an increase in structural anisotropy.

A more revealing picture emerges from the ratio $A_\sigma/A_F$, shown in Fig.~\ref{fig:BC_trend}-(b). This single metric synthesises the results of all tests by expressing the relative strength of mechanical and structural anisotropies in a compact and comparable form. Across all $K_c$ values considered, mechanical anisotropy is found to be approximately $1.75$–$2.5$ times stronger than structural anisotropy, demonstrating a clear hierarchy between the two mechanisms for this material and deposition method. 
Moreover, the gradual decrease of $A_\sigma/A_F$ with increasing $K_c$ shows that the balance between the two mechanisms shifts as the stress state becomes more deviatoric. Although both $A_\sigma$ and $A_F$ grow with $K_c$, the fact that their ratio decreases indicates that the response becomes relatively more sensitive to the fabric than to the stress-induced anisotropy at higher stress ratios. In other words, structural anisotropy, which is associated with the particle arrangement imparted during deposition, does not itself intensify necessarily  with $K_c$, but its relative importance in controlling the initial stress–path inclination grows as the stress state becomes more directionally biased. The ratio $A_\sigma/A_F$ thus provides a concise  measure of this shifting balance between stress-induced and fabric-induced anisotropies, offering a meaningful basis for comparison across materials, stress states, and constitutive models. It is worth noting that, although the specific magnitudes of $A_\sigma$, $A_F$, and their ratio will inevitably depend on the material state (e.~g. void ratio, stress level, fabric intensity), the qualitative trends observed here, namely the hierarchy between the two mechanisms and the increasing relative influence of fabric at higher $K_c$, can be reasonably expected to persist across a broad range of states for granular materials.

\subsection{Comparison with Isotropic Hypoplasticity}

To further examine the separation between mechanical and structural anisotropy and to assess the validity of the first--order linearisation introduced in Eq.~\eqref{eq:linPhi}, the experimental results were compared with predictions from an isotropic hypoplastic model. Hypoplasticity, originally introduced by \cite{Kolymbas1977}, provides a rate--type framework in which the stress rate depends nonlinearly on both the current stress and the strain rate. The sand hypoplasticity model of \cite{Wolffersdorff_1996} incorporates concepts from Critical State Soil Mechanics, including a stress--dependent critical state line and a three--dimensional failure locus based on \cite{Matsuoka1982}. In its general form, the constitutive relation may be written as
\begin{equation}\label{eq:hypo_repeat}
    \dot{\boldsymbol{\sigma}}
    = \hb(\boldsymbol{\sigma}, \dot{\boldsymbol{\varepsilon}}, e)
    = \Lcb \dodo \dot{\boldsymbol{\varepsilon}}
      + \Nb \, \|\dot{\boldsymbol{\varepsilon}}\|,
\end{equation}
where $\Lcb$ and $\Nb$ denote, respectively, the linear and nonlinear parts of the hypoplastic operator, and $e$ is the void ratio. The material parameters used for Fraser River Sand are listed in Table~\ref{tab:sand_hypo}.
These parameters are intended only for a qualitative examination of the model response and should not be interpreted as a validated calibration for Fraser River Sand.

\begin{table}[ht]
\centering
\caption{Parameter set for hypoplasticity~\cite{Wolffersdorff_1996}}
\label{tab:sand_hypo}
\small
\begin{tabular}{|l|l|l|l|l|l|l|l|}
\hline
$\varphi_c$ & $h_s$ (MPa) & $n$ & $e_{d0}$ & $e_{c0}$ & $e_{i0}$ & $\alpha$ & $\beta$ \\
\hline
$34^\circ$ & $2000$ & $0.25$ & $0.5\,e_{c0}$ & $0.9$ & $1.15\,e_{c0}$ & $0.25$ & $1.0$ \\
\hline
\end{tabular}
\end{table}

\begin{figure}
	\centering
\includegraphics[width=\textwidth]{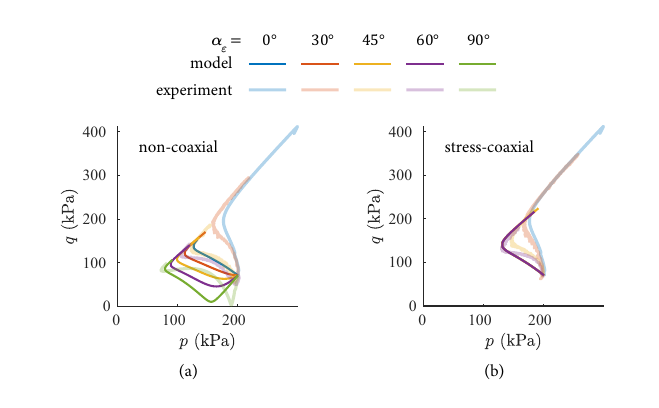}
	\caption{Predictions of an isotropic hypoplasticity for the cases with $K_c=1.5$ and for two test series; (a) non--coaxial, and (b) stress--coaxial. The model predicts directional dependency for the non-coaxial cases, but not for the stress--coaxial series. In the simulations the mean initial void ratios of the experiments are used: (a) $e_\text{ini} = 0.74$, (b) $e_\text{ini} = 0.72$.}
	\label{fig:IsoHypo_predictions}
\end{figure}

Figure~\ref{fig:IsoHypo_predictions} presents the model predictions in comparison with the experimental results for both non--coaxial and stress--coaxial loading configurations.  
For the non--coaxial loading series (Fig.~\ref{fig:IsoHypo_predictions}-(a)), the isotropic hypoplastic model captures a substantial portion of the directional dependence, despite the absence of any explicit structural anisotropy. This behaviour is consistent with the model form in Eq.~\eqref{eq:hypo_repeat}, which inherently incorporates mechanical anisotropy through the stress--dependent cross--invariant $\bs{\sigma}\dodo\dot{\bs{\varepsilon}}$ in the first term. The comparison for the non--coaxial loading is illuminating in that it illustrates the extent to which the directional dependency  originates through the asymmetry of stress tensor alone. Crucially, key qualitative features such as initial load reversal (decrease in $q$) also seem to be captured by the isotropic model. In interpreting these observations it should be recalled that the initial stress is kept coaxial with the initial fabric for  the non--coaxial test series. We expect the accuracy of the predictions to decrease if the preferred orientations of stress and structural anisotropy deviate.

In contrast to the non--coaxial case, the isotropic hypoplastic model is unable to reproduce the directional dependence observed for the stress--coaxial loading series (Fig.~\ref{fig:IsoHypo_predictions}-(b)) and the stress responses for all loading directions are identical. In this configuration, mechanical anisotropy is suppressed by construction ($\Omega_\sigma = 1$) since the strain--rate direction is forced to remain coaxial with the stress, and the observed variation arises entirely from the internal fabric, which is absent from the isotropic model. 

\begin{figure}
	\centering
\includegraphics[width=0.3\textwidth]{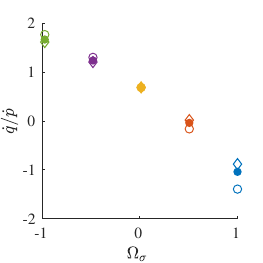}
	\caption{The prediction of the isotropic hypoplastic model for the correlation between the initial slope of stress path and $\Omega_\sigma$. The near linear trend partially confirms the validity of the linearisation approximation in Eq.~\eqref{eq:linPhi} for three different initial void ratios $\circ$  $e_\text{ini}=0.6$, $\bullet$ $e_\text{ini}=0.74$, $\diamond$ $e_\text{ini}=0.85$}
	\label{fig:omega_sigma_hypo}
\end{figure}

The hypoplastic model is also used as an independent check on the linearisation adopted in Eq.~\eqref{eq:linPhi}. Figure~\ref{fig:omega_sigma_hypo} shows the variation of the initial slope with the loading orientation as predicted by the isotropic hypoplastic model. A clear near-linear trend is obtained for multiple initial void ratios, consistent with the first-order linearisation underlying Eq.~\eqref{eq:linPhi}. Given that the model is not predisposed with such a linear dependence, the emergence of this behaviour is noteworthy and provides additional support for the validity of the proposed approximation.

\section{Conclusion}

This paper introduced a simplified yet robust framework for distinguishing mechanical and structural components of anisotropy in granular materials, based on a first--order linearisation of the incremental stress–strain response. The approach expresses the initial stress–path inclination in terms of two cosine--like measures of alignment, $\Omega_\sigma$ and $\Omega_F$, associated with the deviatoric stress and the internal fabric, respectively. Application of this formulation to complementary hollow–cylinder tests allowed the two contributions to be isolated and quantified through the coefficients $A_\sigma$ (mechanical anisotropy) and $A_F$ (structural anisotropy).

The experimental results show that both anisotropy coefficients increase with the initial stress ratio $K_c$, indicating a growing directional sensitivity of the incremental response as the stress state becomes more deviatoric. The rise of $A_\sigma$ is consistent with the expected amplification of stress-induced anisotropy under increasing shear stress, while the increase of $A_F$ reflects a state-dependent enhancement of the influence of fabric. The ratio $A_\sigma/A_F$ synthesises these trends into a single, interpretable metric. Mechanical anisotropy is found to dominate for all stress states examined, yet its relative strength decreases systematically with $K_c$, demonstrating that the fabric becomes comparatively more influential under more deviatoric consolidation.

The proposed decomposition and its experimental quantification provide a practical route for assessing the balance between stress-induced and fabric-induced anisotropies in granular materials. The results offer a physically transparent benchmark for constitutive modelling, including for models that incorporate directional dependence only through cross-invariants of stress and strain rate. More broadly, the framework facilitates comparative studies across materials, deposition methods, and loading paths, and may serve as a basis for future extensions involving evolving fabric or non-linear anisotropy measures.

\section{Acknowledgements}
This research was funded  by the Natural Sciences and Engineering Research Council of Canada (Discovery Grant RGPIN-2020-06480 held by M.~P.) and in part by the Austrian Science Fund (FWF) 10.55776/V918. G.~M. is funded by the FWF. For open access purposes, the authors have applied a CC BY public copyright license to any author accepted manuscript version arising from this submission.

\appendix
\section{Appendix: Linearization of the Incremental Stress Ratio \texorpdfstring{$\dot{q}/\dot{p}$}{dq/dp}}

Starting from the general rate form of the constitutive law and restricting attention to terms that are first-order homogeneous, and incrementally linear in the strain rate, one may write
\begin{equation}
\label{eq:app_rate_general}
\dot{\boldsymbol{\sigma}}
=
\mathbb{C}(\boldsymbol{\sigma},\boldsymbol{F}) : \dot{\boldsymbol{\varepsilon}},
\end{equation}
where $\mathbb{C}$ is an isotropic fourth-order tensor function of the current stress--fabric state. According to the representation theorem of \cite{boehler1979simple,Spencer1971}, such a tensor function may be expanded in tensorial combinations generated by $\boldsymbol{I}$, $\widehat{\boldsymbol{s}}$, and $\widehat{\boldsymbol{F}}$. Retaining only terms that are linear in the anisotropic directions $\widehat{\boldsymbol{s}}$ and $\widehat{\boldsymbol{F}}$ gives the first-order form
\begin{equation}
\label{eq:app_sigma_rep}
\begin{aligned}
\dot{\boldsymbol{\sigma}}
={}&
K_0\,\dot{\varepsilon}_v\,\boldsymbol{I}
+2G_0\,\dot{\boldsymbol{e}}
+K_{\sigma}\,(\widehat{\boldsymbol{s}}:\dot{\boldsymbol{e}})\,\boldsymbol{I}
+K_F\,(\widehat{\boldsymbol{F}}:\dot{\boldsymbol{e}})\,\boldsymbol{I}
\\
&\quad
+G_{\sigma v}\,\dot{\varepsilon}_v\,\widehat{\boldsymbol{s}}
+G_{Fv}\,\dot{\varepsilon}_v\,\widehat{\boldsymbol{F}}
+H_{\sigma}\,\operatorname{dev}\!\left(
\widehat{\boldsymbol{s}}\,\dot{\boldsymbol{e}}
+\dot{\boldsymbol{e}}\,\widehat{\boldsymbol{s}}
\right)
+H_F\,\operatorname{dev}\!\left(
\widehat{\boldsymbol{F}}\,\dot{\boldsymbol{e}}
+\dot{\boldsymbol{e}}\,\widehat{\boldsymbol{F}}
\right),
\end{aligned}
\end{equation}
where the scalar coefficients depend only on the current state through scalar invariants, and the operator \(\operatorname{dev}(\bm{X})\) gives the deviatoric part of the second-order tensor \(\bm{X}\). Higher-order terms are neglected.
Taking the trace of this expression gives
\begin{equation}
\label{eq:app_pdot}
\dot{p}
=
K_0\,\dot{\varepsilon}_v
+
K_{\sigma}\,(\widehat{\boldsymbol{s}}:\dot{\boldsymbol{e}})
+
K_F\,(\widehat{\boldsymbol{F}}:\dot{\boldsymbol{e}}).
\end{equation}
Introducing \(r= {\dot{\varepsilon}_v}/{\lVert \dot{\boldsymbol{e}} \rVert}\)
and using $\Omega_{\sigma}$ and $\Omega_F$ as defined in Eq.~\eqref{eq:Omegas}, this becomes
\begin{equation}
\label{eq:app_pdot_reduced}
\dot{p}
=
\lVert \dot{\boldsymbol{e}} \rVert
\left(
K_0\,r
+
K_{\sigma}\,\Omega_{\sigma}
+
K_F\,\Omega_F
\right).
\end{equation}

For the deviatoric stress magnitude we can write \(\dot{q}=\sqrt{\frac{3}{2}}\,
\widehat{\boldsymbol{s}}:\dot{\boldsymbol{\sigma}}\).
Substituting the first-order approximation form Eq.~\eqref{eq:app_sigma_rep} and collecting the resulting scalar projections gives
\begin{equation}
\label{eq:app_qdot_full}
\dot{q}
=
\lVert \dot{\boldsymbol{e}} \rVert
\left(
\widetilde{M}_v\,r
+
\widetilde{M}_{\sigma}\,\Omega_{\sigma}
+
\widetilde{M}_1\,\Theta_1
+
\widetilde{M}_2\,\Theta_2
\right),
\end{equation}
where
\begin{equation}
\label{eq:app_theta_defs}
\Theta_1
=
\operatorname{dev}\!\left(\widehat{\boldsymbol{s}}^{\,2}\right):\widehat{\dot{\boldsymbol{e}}},
\qquad
\Theta_2
=
\operatorname{dev}\!\left(
\widehat{\boldsymbol{s}}\,\widehat{\boldsymbol{F}}
+\widehat{\boldsymbol{F}}\,\widehat{\boldsymbol{s}}
\right):\widehat{\dot{\boldsymbol{e}}}.
\end{equation}
From a physical perspective,  the anisotropy of the state is assumed to be fully characterised by the stress and fabric directions, and therefore, higher-order combinations of these tensors do not introduce new independent anisotropy directions. Mathematically, this implies that the higher-order tensor combinations appearing in Eq.~\eqref{eq:app_theta_defs} remain within the space spanned by them, i.~e.  \(\operatorname{dev}(\widehat{\boldsymbol{s}}^{\,2}),\ \operatorname{dev}\!\left(\widehat{\boldsymbol{s}}\,\widehat{\boldsymbol{F}}+\widehat{\boldsymbol{F}}\,\widehat{\boldsymbol{s}}\right)\in \operatorname{span}\{\widehat{\boldsymbol{s}},\widehat{\boldsymbol{F}}\}\). 
Under this condition, the variables $\Theta_1$ and $\Theta_2$ can be expressed as a linear combination of the two scalar coordinates $\Omega_{\sigma}$ and $\Omega_F$, as
\begin{equation}
\label{eq:app_theta_reduction}
\Theta_1
=
c_{1\sigma}\,\Omega_{\sigma}
+
c_{1F}\,\Omega_F,
\qquad
\Theta_2
=
c_{2\sigma}\,\Omega_{\sigma}
+
c_{2F}\,\Omega_F,
\end{equation}
where the four $c$ coefficients  depend only on the current stress--fabric state. Substituting Eq.~\eqref{eq:app_theta_reduction} into \eqref{eq:app_qdot_full}, the projected deviatoric rate may therefore be written in the reduced form
\begin{equation}
\label{eq:app_qdot_reduced}
\dot{q}
=
\lVert \dot{\boldsymbol{e}} \rVert
\left(
M_v\,r
+
M_{\sigma}\,\Omega_{\sigma}
+
M_F\,\Omega_F
\right),
\end{equation}
where the redefined coefficients again depend only on the current state.
The incremental stress-path slope then becomes
\begin{equation}
\label{eq:app_ratio_general}
\frac{\dot{q}}{\dot{p}}
=
\frac{
M_v\,r
+
M_{\sigma}\,\Omega_{\sigma}
+
M_F\,\Omega_F
}{
K_0\,r
+
K_{\sigma}\,\Omega_{\sigma}
+
K_F\,\Omega_F
}.
\end{equation}

The local expansion is taken about the aligned (coaxial) reference state, \(\Omega_{\sigma}= \Omega_F=1\).
A first-order Taylor expansion about this point gives
\begin{equation}
\label{eq:app_ratio_shifted}
\frac{\dot{q}}{\dot{p}}
\approx
\eta_{\ast}
+
A_{\sigma}\,(\Omega_{\sigma}-1)
+
A_F\,(\Omega_F-1),
\end{equation}
with
\begin{equation}
\label{eq:app_eta_star}
\begin{gathered}
\eta_{\ast}
=
\frac{N_{\ast}}{D_{\ast}},  \qquad  A_{\sigma}
=
\frac{M_{\sigma}\,D_{\ast}-N_{\ast}\,K_{\sigma}}{D_{\ast}^{2}},
\qquad
A_F
=
\frac{M_F\,D_{\ast}-N_{\ast}\,K_F}{D_{\ast}^{2}},\\
N_{\ast}
=
M_v\,r + M_{\sigma} + M_F,
\qquad
D_{\ast}
=
K_0\,r + K_{\sigma} + K_F.
\end{gathered}
\end{equation}
This may be recast in the affine form postulated in Eq.~\eqref{eq:linPhi},
\begin{equation}
\label{eq:app_ratio_affine}
\frac{\dot{q}}{\dot{p}}
\approx
\eta_0
+
A_{\sigma}\,\Omega_{\sigma}
+
A_F\,\Omega_F,
\end{equation}
where \(\eta_0= \eta_{\ast}-A_{\sigma}-A_F\).
For the undrained tests considered in this study, one sets $r=0$ in the preceding expressions.

\bibliographystyle{ieeetr}
\bibliography{refs}%

@article{sivathayalan2002influence,
	title={Influence of generalized initial state and principal stress rotation on the undrained response of sands},
	author={Sivathayalan, S and Vaid, Y P},
	journal={Canadian Geotechnical Journal},
	volume={39},
	number={1},
	pages={63--76},
	year={2002},
	publisher={NRC Research Press Ottawa, Canada}
}

@Article{Kolymbas1977,
  author    = {Kolymbas, D.},
  title     = {A rate-dependent constitutive equation for soils},
  journal   = {Mechanics Research Communications},
  year      = {1977},
  volume    = {4},
  pages     = {367-372},
  doi       = {10.1016/0093-6413(77)90056-8}
}

@InProceedings{Matsuoka1982,
  author    = {H. Matsuoka and T. Nakai},
  title     = {A new failure criterion for soils in three dimensional stress},
  booktitle = {IUTAM Conference on deformation and failure of granular materials},
  year      = {1982},
  pages     = {253--263},
  address   = {Delft},
}

@article{Wolffersdorff_1996,
  author =        {von Wolffersdorff, P.-A.},
  journal =       {Mechanics of Cohesive-Frictional Materials, 1},
  pages =         {251-271},
  title =         {A hypoplastic relation for granular materials with a
                   predefined limit state surface},
  volume =        {1},
  year =          {1996},
  doi =
  {10.1002/(SICI)1099-1484(199607)1:3<251::AID-CFM13>3.0.CO;2-3},
}

@article{prasanna2020effects,
	title={Effects of initial direction and subsequent rotation of principal stresses on liquefaction potential of loose sand},
	author={Prasanna, R and Sinthujan, N and Sivathayalan, Siva},
	journal={Journal of Geotechnical and Geoenvironmental Engineering},
	volume={146},
	number={3},
	pages={04019130},
	year={2020},
	publisher={American Society of Civil Engineers}
}

@article{zhao2013unique,
  title={Unique critical state characteristics in granular media considering fabric anisotropy},
  author={Zhao, Jidong and Guo, Ning},
  journal={G{\'e}otechnique},
  volume={63},
  number={8},
  pages={695--704},
  year={2013},
  publisher={Thomas Telford Ltd}
}

@article{wong1985induced,
  title={Induced and inherent anisotropy in sand},
  author={Wong, RKS and Arthur, JRF},
  journal={Geotechnique},
  volume={35},
  number={4},
  pages={471--481},
  year={1985},
  publisher={Thomas Telford Ltd}
}

@article{wiebicke2020measuring,
  title={Measuring the evolution of contact fabric in shear bands with X-ray tomography},
  author={Wiebicke, Max and And{\`o}, Edward and Viggiani, Gioacchino and Herle, Ivo},
  journal={Acta Geotechnica},
  volume={15},
  number={1},
  pages={79--93},
  year={2020},
  publisher={Springer}
}

@article{oda1993inherent,
  title={Inherent and induced anisotropy in plasticity theory of granular soils},
  author={Oda, M},
  journal={Mechanics of Materials},
  volume={16},
  number={1-2},
  pages={35--45},
  year={1993},
  publisher={Elsevier}
}

@article{premnath2023effects,
  title={Effects of principal strain direction and intermediate principal strain on undrained shear behavior of sand},
  author={Premnath, S and Pouragha, M and Prasanna, R and Sivathayalan, S},
  journal={Journal of Geotechnical and Geoenvironmental Engineering},
  volume={149},
  number={7},
  pages={04023048},
  year={2023},
  publisher={American Society of Civil Engineers}
}

@article{arthur1980principal,
  title={Principal stress rotation: a missing parameter},
  author={Arthur, J Robin F and Rodriguez del C, Juan I and Dunstan, Treve and Chua, Ken S},
  journal={Journal of the Geotechnical Engineering Division},
  volume={106},
  number={4},
  pages={419--433},
  year={1980},
  publisher={American Society of Civil Engineers}
}

@article{symes1984undrained,
  title={Undrained anisotropy and principal stress rotation in saturated sand},
  author={Symes, MJPR and Gens, A and Hight, DW},
  journal={Geotechnique},
  volume={34},
  number={1},
  pages={11--27},
  year={1984},
  publisher={Thomas Telford Ltd}
}

@article{rothenburg1989analytical,
  title={Analytical study of induced anisotropy in idealized granular materials},
  author={Rothenburg, Leo and Bathurst, RJ},
  journal={Geotechnique},
  volume={39},
  number={4},
  pages={601--614},
  year={1989},
  publisher={Thomas Telford Ltd}
}

@article{boehler1979simple,
  title={A simple derivation of representations for non-polynomial constitutive equations in some cases of anisotropy},
  author={Boehler, Jean-Paul},
  journal={ZAMM-Journal of Applied Mathematics and Mechanics/Zeitschrift f{\"u}r Angewandte Mathematik und Mechanik},
  volume={59},
  number={4},
  pages={157--167},
  year={1979},
  publisher={Wiley Online Library}
}

@incollection{Spencer1971,
  author    = {Spencer, A. J. M.},
  title     = {Theory of Invariants},
  booktitle = {Continuum Physics, Volume I},
  editor    = {Eringen, A. C.},
  publisher = {Academic Press},
  address   = {New York},
  year      = {1971},
  pages     = {239--353},
  doi       = {10.1016/B978-0-12-240801-4.50008-X}
}

@article{yang2008quantifying,
  title={Quantifying and modelling fabric anisotropy of granular soils},
  author={Yang, ZX and Li, XS and Yang, J},
  journal={G{\'e}otechnique},
  volume={58},
  number={4},
  pages={237--248},
  year={2008},
  publisher={Thomas Telford Ltd}
}

@article{imseeh20183d,
  title={3D experimental quantification of fabric and fabric evolution of sheared granular materials using synchrotron micro-computed tomography},
  author={Imseeh, Wadi H and Druckrey, Andrew M and Alshibli, Khalid A},
  journal={Granular Matter},
  volume={20},
  number={2},
  pages={24},
  year={2018},
  publisher={Springer}
}

@article{oda1985stress,
  title={Stress-induced anisotropy in granular masses},
  author={Oda, Masanobu and Nemat-Nasser, Siavouche and Konishi, Junichi},
  journal={Soils and foundations},
  volume={25},
  number={3},
  pages={85--97},
  year={1985},
  publisher={The Japanese Geotechnical Society}
}

@article{majmudar2005contact,
  title={Contact force measurements and stress-induced anisotropy in granular materials},
  author={Majmudar, Trushant S and Behringer, Robert P},
  journal={nature},
  volume={435},
  number={7045},
  pages={1079--1082},
  year={2005},
  publisher={Nature Publishing Group UK London}
}

@article{gu2024quantifying,
  title={Quantifying fabric anisotropy of granular materials using wave velocity anisotropy: a numerical investigation},
  author={Gu, Xiaoqiang and Liang, Xiaomin and Hu, Jing},
  journal={G{\'e}otechnique},
  volume={74},
  number={12},
  pages={1263--1275},
  year={2024},
  publisher={Emerald Publishing Limited}
}

@article{kuwano2000anisotropic,
  title={Anisotropic stiffness measurements in a stress-path triaxial cell},
  author={Kuwano, R and Connolly, TM and Jardine, RJ},
  journal={Geotechnical Testing Journal},
  volume={23},
  number={2},
  pages={141--157},
  year={2000},
  publisher={ASTM International}
}

@article{li2012anisotropic,
  title={Anisotropic critical state theory: role of fabric},
  author={Li, Xiang Song and Dafalias, Yannis F},
  journal={Journal of engineering mechanics},
  volume={138},
  number={3},
  pages={263--275},
  year={2012},
  publisher={American Society of Civil Engineers}
}

@article{pouragha2020micromechanical,
  title={Micromechanical correlation between elasticity and strength characteristics of anisotropic rocks},
  author={Pouragha, Mehdi and Eghbalian, Mahdad and Wan, Richard},
  journal={International Journal of Rock Mechanics and Mining Sciences},
  volume={125},
  pages={104154},
  year={2020},
  publisher={Elsevier}
}

@article{fakharian2022influences,
  title={Influences of initial anisotropy and principal stress rotation on the undrained monotonic behavior of a loose silica sand},
  author={Fakharian, Kazem and Kaviani-Hamedani, Farzad and Imam, SM Reza},
  journal={Canadian Geotechnical Journal},
  volume={59},
  number={6},
  pages={847--862},
  year={2022},
  publisher={Canadian Science Publishing 1840 Woodward Drive, Suite 1, Ottawa, ON K2C 0P7}
}

@article{wijewickreme2008experimental,
  title={Experimental observations on the response of loose sand under simultaneous increase in stress ratio and rotation of principal stresses},
  author={Wijewickreme, Dharma and Vaid, Yoginder P},
  journal={Canadian Geotechnical Journal},
  volume={45},
  number={5},
  pages={597--610},
  year={2008}
}

@article{wijewickreme1993behaviour,
  title={Behaviour of loose sand under simultaneous increase in stress ratio and principal stress rotation},
  author={Wijewickreme, Dharmapriya and Vaid, Yoginder P},
  journal={Canadian Geotechnical Journal},
  volume={30},
  number={6},
  pages={953--964},
  year={1993},
  publisher={NRC Research Press Ottawa, Canada}
}

\end{document}